\documentclass[11pt,twoside,letterpaper]{article} 
\usepackage{times,fancyhdr}
\usepackage{graphicx}
\usepackage{amssymb}
\usepackage{amsmath}
\usepackage{amsfonts}

\def\1{{\bf 1}}

\def\be{\begin{equation}}
\def\ee{\end{equation}}


\makeatletter
\def\cleardoublepage{\clearpage\if@twoside \ifodd\c@page\else%
    \hbox{}%
    \thispagestyle{empty}%
    \newpage%
    \if@twocolumn\hbox{}\newpage\fi\fi\fi}
\makeatother

\def\figurename{Figure}
\makeatletter
\renewcommand{\fnum@figure}[1]{\figurename~\thefigure.}
\makeatother

\def\tablename{Table}
\makeatletter
\renewcommand{\fnum@table}[1]{\tablename~\thetable.}
\makeatother

\begin{document}
\title{
{\begin{flushleft}
\vskip 0.45in
{\normalsize\bfseries\textit{Chapter~1}}
\end{flushleft}
\vskip 0.45in
\bfseries\scshape The Nature of the Infectious Agents: PrP Models of Resistant Species to Prion Diseases (Dog, Rabbit and Horses)}}
\author{\bfseries\itshape Jiapu Zhang\thanks{E-mail addresses: jiapu\_zhang@hotmail.com, j.zhang@ballarat.edu.au, Phone: 61-423 487 360}\\
Centre for Informatics and Applied Optimization \&\\
Graduate School of Sciences, Information Technology and Engineering,\\
The University of Ballarat, MT Helen Campus, Victoria 3353, Australia
}
\date{}
\maketitle
\thispagestyle{empty}
\setcounter{page}{1}
\thispagestyle{fancy}
\fancyhead{}
\fancyhead[L]{In: Prions and Prion Diseases: New Developments \\
Editor: J. M. Verdier, pp. {\thepage-\pageref{lastpage-01}}} 
\fancyhead[R]{ISBN 1621000273  \\
\copyright~2012 Nova Science Publishers, Inc.}
\fancyfoot{}
\renewcommand{\headrulewidth}{0pt}
%




\begin{abstract}
Prion diseases are invariably fatal and highly infectious neurodegenerative diseases affecting humans and animals. The neurodegenerative diseases such as Creutzfeldt-Jakob disease, variant Creutzfeldt-Jakob diseases, Gerstmann-Str$\ddot{\text{a}}$ussler-Scheinker syndrome, Fatal Familial Insomnia, Kuru in humans, scrapie in sheep, bovine spongiform encephalopathy (or 'mad-cow' disease) and chronic wasting disease in cattle etc belong to prion diseases. By now there have not been some effective therapeutic approaches to treat all these prion diseases. In 2008, canine mammals including dogs (canis familials) were the first time academically reported to be resistant to prion diseases (Vaccine 26: 2601--2614 (2008)). Rabbits are the mammalian species known to be resistant to infection from prion diseases from other species (Journal of Virology 77: 2003–-2009 (2003)). In 2010 horses were reported to be resistant to prion diseases too (Proceedings of the National Academy of Sciences USA 107: 19808--19813 (2010)). By now all the NMR structures of dog, rabbit and horse prion proteins had been released into protein data bank respectively in 2005, 2007 and 2010 (Proceedings of the National Academy of Sciences USA 102: 640--645 (2005), Journal of Biomolecular NMR 38:181 (2007), Journal of Molecular Biology 400: 121-–128 (2010)). Thus, at this moment it is very worth studying the NMR molecular structures of horse, dog and rabbit prion proteins to obtain insights into their immunity prion diseases.

This article reports the findings of the molecular structural dynamics of wild-type horse, dog, and rabbit prion proteins. The dog and horse prion proteins have stable molecular structures whether under neutral or low pH environments. Rabbit prion protein has been found having stable molecular structures under neutral pH environment, but without structural stability  under low pH environment. Under low pH environment, the salt bridges such as D177--R163 were broken and caused the collapse of the stable $\alpha$-helical molecular structures. The contribution of D177--R163 salt bridge was confirmed by the recent experimental results on the X-ray and NMR structures of rabbit prion protein (Proceedings of National Academy of Sciences USA 107: 19808–-19813 (2010), Journal of Biological Chemistry 285 (41), 31682--31693 (2010), Journal of Molecular Biology 400: 121–-128 (2010)). This salt bridge might have a value to the scientific community in its drive to find treatments for prion diseases.
\end{abstract}

\noindent \textbf{Keywords:} Prion diseases; Immunity; Rabbit, dog, and horse prion proteins; Molecular dynamics.

\pagestyle{fancy}
\fancyhead{}
\fancyhead[EC]{Jiapu Zhang}
\fancyhead[EL,OR]{\thepage}
\fancyhead[OC]{The Nature of the Infectious Agents}
\fancyfoot{}
\renewcommand\headrulewidth{0.5pt}

\section{Introduction}
Prion diseases, including Creutzfeldt-Jakob disease (CJD), variant Creutzfeldt-Jakob diseases (vCJD), Gerstmann-Str$\ddot{\text{a}}$ussler-Scheinker syndrome (GSS), Fatal Familial Insomnia (FFI), Kuru in humans, scrapie in sheep, bovine spongiform encephalopathy (BSE or `mad-cow' disease) and chronic wasting disease (CWD) in cattle, etc., are invariably fatal and highly infectious neurodegenerative diseases affecting humans and animals. However, for treating all these diseases, there is no effective therapeutic approach \cite{aguzzi_etal, prusiner, weissmann}.

Fortunately, dogs, rabbits, and horses have been reported to be the few mammalian species being resistant to infection from prion diseases. Their NMR molecular structures were released into the Protein Data Bank \cite{pdb_bank} in the years 2003, 2005, 2010 respectively. The N-terminal residues of prion proteins are unstructured and not included in this study. The C-terminal residues are well structured, with 3 $\alpha$-helices, 2 short anti-parallel $\beta$-sheets, and a disulfide bond between the 2nd and 3rd $\alpha$-helix (residues number 178 and number 213). The infectious prion (PrP$^{\text{Sc}}$) is an abnormally folded form of the normal cellular prion (PrP$^{\text{C}}$) and the conversion of PrP$^{\text{C}}$ to PrP$^{\text{Sc}}$ is believed to involve conformational change from a predominantly $\alpha$-helical protein (42\% $\alpha$-helix, 3\% $\beta$-sheet) to a protein rich in $\beta$-sheets (30\% $\alpha$-helix, 43\% $\beta$-sheet).

Rabbits are one of the few mammalian species reported to be resistant to infection from prion diseases isolated from other species \cite{vorberg_etal}. Recently, the NMR molecular structures of wild-type, mutant S173N and mutant I214V rabbit prion proteins (124-228) were released into the Protein Data Bank \cite{pdb_bank} with PDB ID codes 2FJ3, 2JOH, 2JOM respectively. Zhang (2010a) studied these NMR structures by MD simulations and simulation results at {\sl 450 K} under low and neutral pH environments confirmed the structural stability of wild-type rabbit prion protein. However, the previous MD simulation results at {\sl 500 K} under neutral pH environment showed that wild-type rabbit prion protein (124-228) does not have a structural stability \cite{zhang2009}. At {\sl 450 K}, Zhang (2011a) also studied the molecular dynamics of NMR structures of human and mouse prion proteins (1QLX.pdb and 1AG2.pdb respectively) to make comparisons with that of rabbit prion protein \cite{zhang2011a}. The comparisons show that the enhanced stability of the C-terminal ordered region especially helix 2 through the D177–-R163 salt-bridge formation renders the rabbit prion protein stable, the salt bridge D201–-R155 linking helices 3 and 1 and the hydrogen bond H186–-R155 also contribute to the structural stability of rabbit prion protein; however, the human and mouse prion protein structures were not affected by the removing these two salt bridges.

Dogs (canis familials, canine mammals) were the first time academically reported in 2008 to be resistant to prion diseases \cite{polymenidoua2008}. Zhang et al. (2011) studied the molecular dynamics of dog prion protein (1XYK.pdb) at {\sl 450 K} and found that the dog prion protein has stable molecular structures whether under neutral or low pH environments \cite{zhang2011b}.

Horses were academically reported to be resistant to prion diseases recently \cite{perez2010}. Zhang (2011) studied the molecular dynamics of horse prion protein (2KU4.pdb) at {\sl 450 K} and found that the dog prion protein has stable molecular structures whether under neutral or low pH environments \cite{zhang2011c}. Molecular dynamics results show that the strong salt bridge ASP177--ARG163 contributes to the structural stability of horse prion protein \cite{zhang2011c}.

All the above molecular dynamics simulations done for rabbit, dog and horse prion proteins are at {\sl 450 K} and {\sl 300 K} under neutral and low pH environments. The simulations done at room temperature {\sl 300 K} whether under neutral or low pH environment display very little fluctuation for rabbit, dog, horse, human and mouse prion proteins. The common result for rabbit, dog and horse prion proteins found at {\sl 450 K} is: there is a strong bridge ASP177--ARG163 (like a taut bow string) keeping the $\beta$2-$\alpha$2 loop of prion proteins linked. The finding agrees with the recent experimental results on the $\beta$2-$\alpha$2 loop \cite{bett2012, linw2011, fernandez-funez2011, perez2010, khan2010, sweeting2009, wen2010a, wen2010b}, which plays an important role to stabilize the structural stability of prion proteins.

This Chapter reports the molecular dynamics simulation results done for rabbit, dog and horse prion proteins are at {\sl 350 K} under neutral and low pH environments. Molecular dynamics results agree with the results found at {\sl 450 K}. The stable $\alpha$-helical secondary structure of rabbit prion protein collapses under low pH environment. The strong bridge ASP177--ARG163 (like a taut bow string) keeping the $\beta$2-$\alpha$2 loop of prion proteins linked contributes to the structural stability of rabbit, dog and horse prion proteins. This salt bridge might be a potential drug target and have a value to the scientific community in its drive to find treatments for prion diseases. The rest of this article is arranged as follows. The molecular dynamics simulation materials and methods are introduced in Section 2. Section 3 mainly gives molecular dynamics simulation results and their discussions. Concluding remarks are given in the last Section.

\section{Materials and Methods}
The molecular dynamics simulation materials and methods for horse, dog and rabbit prion proteins are completely same as the ones of \cite{zhang2010}. Simulation initial structure for the dog and rabbit prion proteins were built on HoPrP$^{\text{C}}$(119-231) (PDB entry 2KU4), DoPrP$^{\text{C}}$(121-231) (PDB entry 1XYK) and RaPrP$^{\text{C}}$(124–-228) (PDB entry 2FJ3), respectively. Simulations were done under low pH and normal pH environments respectively. All the simulations were performed with the AMBER 11 package \cite{amber11}, with analysis carried out using functionalities in AMBER 9 \cite{amber9} and AMBER 7 CARNAL \cite{amber7}. Graphs were drawn by XMGRACE of Grace 5.1.21, DSSP \cite{dssp}.

All simulations used the ff03 force field of the AMBER 11 package, in neutral and low pH environments (where residues HIS, ASP, GLU were changed into HIP, ASH, GLH respectively by the XLEaP module of AMBER 11 in order to get the low pH environment). The systems were surrounded with a 12 angstrom layer of TIP3PBOX water molecules and neutralized by sodium ions using XLEaP module of AMBER 11. The solvated proteins with their counterions were minimized mainly by the steepest descent method and then a small number of conjugate gradient steps were performed on the data, in order to remove bad hydrogen bond contacts. Then the solvated proteins were heated from {\sl 100} to {\sl 300 K} during 1 ns (with step size 1 fs) and from {\sl 300} to {\sl 350 K} during 1 ns (with step size 2 fs). The thermostat algorithm used is the Langevin thermostat algorithm in constant NVT ensembles. The SHAKE algorithm and PMEMD algorithm with nonbonded cutoffs of 12 angstrom were used during the heating. Equilibrations were done in constant NPT ensembles under Langevin thermostat for 2 ns. After equilibrations, production MD phase was carried out at {\sl 350 K} for 30 ns using constant pressure and temperature ensemble and the PMEMD algorithm with nonbonded cutoffs of 12 angstrom during simulations. Step size for the production runs is 2 fs. The structures were saved to file every 1000 steps.

\section{Results and Discussion}
\begin{figure}[h!]
\centerline{
\includegraphics[width=2.64in]{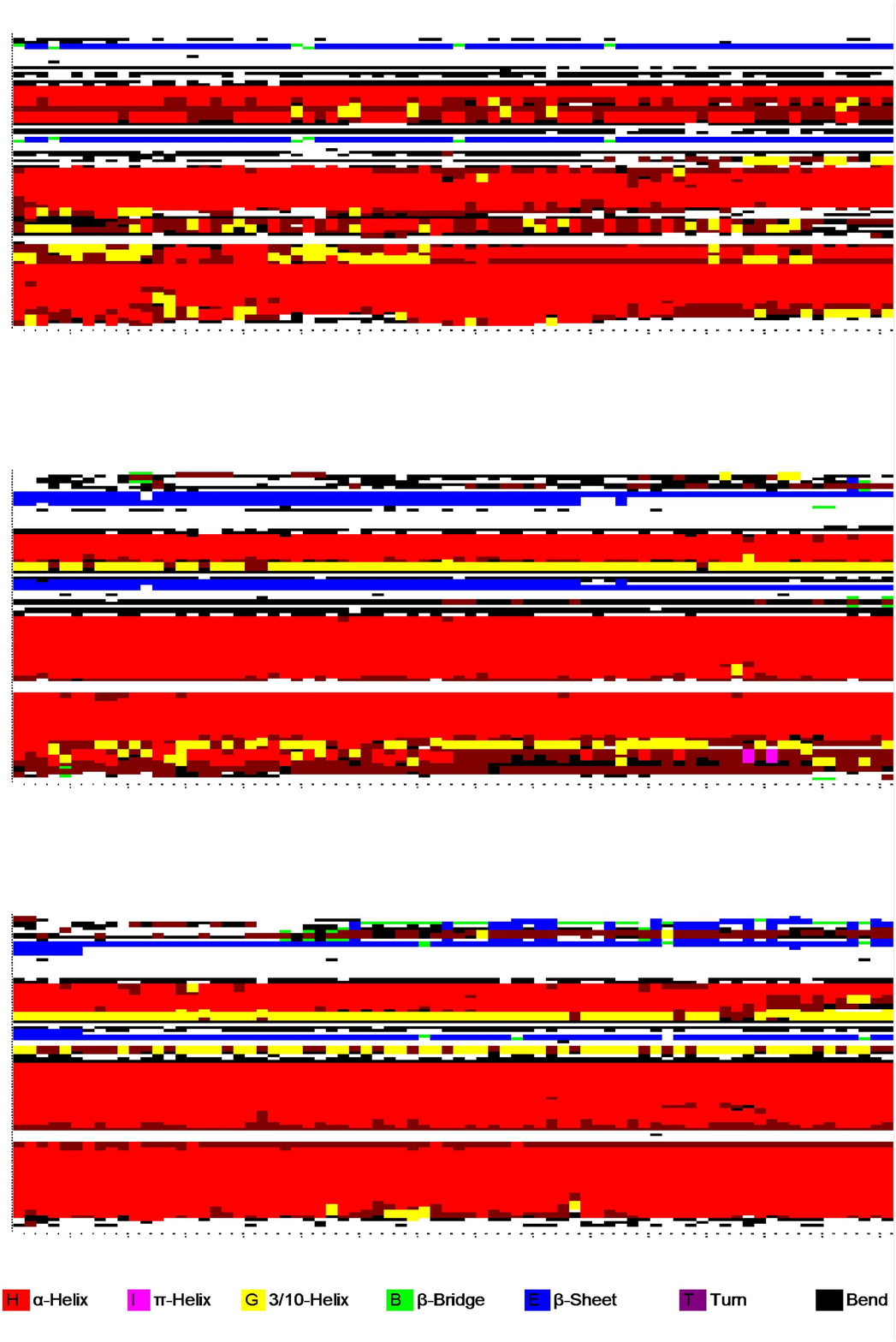} \quad
\includegraphics[width=2.64in]{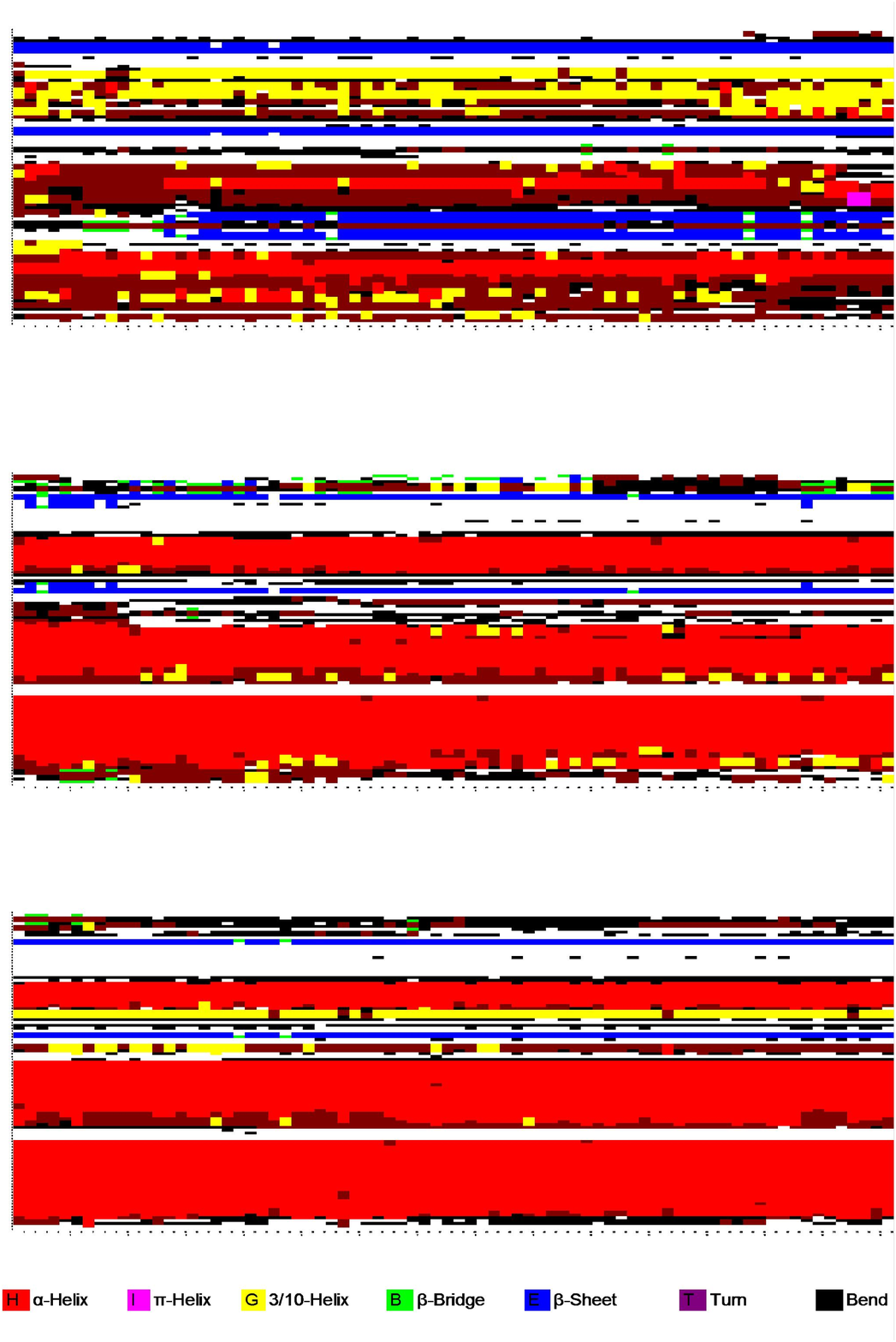}
}
\caption{Secondary structures of rabbit, dog and horse prion proteins (from up to down) at {\it 350 K} under neutral to low pH
environments (from left to right) (X-axis: 0 ns - 30 ns (from left to right), Y-axis: residue numbers 124 - 228 / 121 - 231 / 119 - 231 (from up to down)).}
\label{secondary_structures}
\end{figure}

At {\it 350 K} there are fluctuation and variation among rabbit, dog and horse prion proteins, but their backbone atom RMSDs (root mean square deviations) respectively calculated from their minimized structures and their radii of gyrations do not have great difference even under low pH environment. Their secondary structures under neutral pH environment at {\it 350 K} do not change very much either. However, the secondary structures under low pH environment at {\it 350 K} have great differences between rabbit prion protein and dog and horse prion proteins (Figure \ref{secondary_structures}): the $\alpha$-helices of rabbit prion protein were completely unfolded and began to turn into $\beta$-sheets but those of dog and horse prion proteins were not changed very much. These results indicate the C-terminal region of RaPrP$^{\text{C}}$ has lower thermostability than that of DoPrP$^{\text{C}}$ and HoPrP$^{\text{C}}$. Under the low pH environment, the important salt bridges such as D177-R163, D201-R155 were removed (thus the free energies of the salt bridges changed the thermostability) so that the structure nearby the central helices 1-3 was changed for rabbit prion protein.

\begin{figure}[h!]
\centerline{
\includegraphics[width=5.24in]{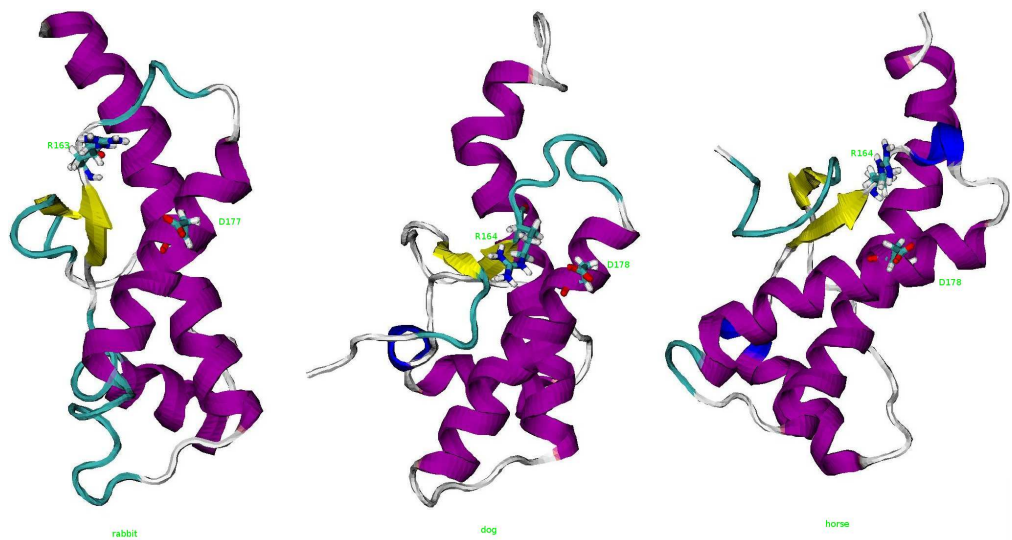}
}
\caption{The salt bridge D177-R163 of rabbit prion protein, D178-R164 of dog and horse prion proteins.}
\label{common_salt_bridge}
\end{figure}

There always exists a strong salt bridge (where the oxygen-nitrogen distance cut-off calculated for the salt bridges is 3.2 angstroms) between D177-R163 for rabbit prion protein, and D178-R164 for dog and horse prion proteins (Figure \ref{common_salt_bridge})
under neutral pH environment at {\it 300 K}, {\it 350 K}, and {\it 450 K}. We can see in Figure \ref{common_salt_bridge} that this salt bridge is just like a taut bow string  linking $\beta$2-sheet and $\alpha$2-helix. It contributes to the structural stability of all these prion proteins. Thus, the region of $\beta$2--$\alpha$2 loop should be a potential drug target region according to recent reports on the $\beta$2--$\alpha$2 loop of prion proteins \cite{bett2012, linw2011, fernandez-funez2011, perez2010, khan2010, sweeting2009, wen2010a, wen2010b}.

Lastly, we compare rabbit with dog and horse prion proteins through the pairwise sequence and structure alignments. We find that there are more identities and similarities between horse and dog prion proteins, compared with the identities and similarities between dog and rabbit prion proteins, and between horse and rabbit prion proteins. This point can be furthermore confirmed by the following data of the Needleman-Wunsch or Smith-Waterman Pairwise Sequence Alignment and the jCE algorithm or jCE Circular Permutation Pairwise Structure Alignment. (1) For dogs and horses, the Identities are 89.38 \% (query) and 90.99 \% (subject) and the Similars are 93.81 \% (query) and 95.50 \% (subject) for the Sequence Alignment, and the Identity is 89.11\% and the Similarity is 93.07\% for the Structure Alignment. (2) For dogs and rabbits, the Identities are  90.09 \% (query) and  72.46 \% (subject) and the Similars are  95.95 \% (query) and  77.18 \% (subject) for the Sequence Alignment, and the Identity is 92.93 \% and the Similarity is 98.99 \% for the Structure Alignment. (3) For horses and rabbits, the Identities are 87.61 \% (query) and 71.74 \% (subject) and the Similars are 93.81 \% (query) and 76.81 \% (subject) for the Sequence Alignment, and the Identity is 71.72\% and the Similarity is 77.78\% for the Structure Alignment. This might explain the reasons why rabbit prion protein differs very much from dog and horse prion proteins in secondary structures under low pH environment \cite{zhang2011b, zhang2011c}.

\section{Conclusion}
The paper is a straight forward molecular dynamics simulation study of rabbit, dog and horse prion proteins which apparently resist the formation of the scrapie form. The analyses of molecular dynamics results confirmed the structural stability of rabbit prion protein under neutral pH environment, and of dog and horse prion proteins under neutral and low pH environments. The main point is that the enhanced stability of the C-terminal ordered region especially helix 2 through the D177-R163 (D178-R164) salt-bridge formation renders the rabbit, dog and horse prion proteins stable. This salt bridge might be a potential drug target for prion diseases. Zhang (2011) systemically analyzes this salt bridge in \cite{zhang2011d} and, like 'GN8' of \cite{kuwara2007}, fixing the distance between D177 and R163 to form a salt bridge distance might be able to design a drug for treating prion diseases. 

\section*{Acknowledgments} This research has been supported by a Victorian Life Sciences Computation Initiative (VLSCI) grant number VR0063 on its Peak Computing Facility at the University of Melbourne, an initiative of the Victorian Government.

\label{lastpage-01}

\end{document}